\documentclass[aps,twocolumn,showpacs,superscriptaddress,floatfix]{revtex4}
\usepackage{graphicx}
\usepackage{psfrag}
\begin{document}
\newcommand{\beq}{\begin{equation}}
\newcommand{\eeq}{\end{equation}}
\newcommand{\ben}{\begin{eqnarray}}
\newcommand{\een}{\end{eqnarray}}
\newcommand{\bea}{\begin{array}}
\newcommand{\eea}{\end{array}}
\newcommand{\om}{(\omega )}
\newcommand{\bef}{\begin{figure}}
\newcommand{\eef}{\end{figure}}
\newcommand{\leg}[1]{\caption{\protect\rm{\protect\footnotesize{#1}}}}
\newcommand{\ew}[1]{\langle{#1}\rangle}
\newcommand{\be}[1]{\mid\!{#1}\!\mid}
\newcommand{\no}{\nonumber}
\newcommand{\etal}{{\em et~al }}
\newcommand{\geff}{g_{\mbox{\it{\scriptsize{eff}}}}}
\newcommand{\da}[1]{{#1}^\dagger}
\newcommand{\cf}{{\it cf.\/}\ }
\newcommand{\ie}{{\it i.e.\/}\ }

\title{Topological Non--connectivity Threshold in long-range spin systems}

\author{F.~Borgonovi}
\affiliation{Dipartimento di
Matematica e Fisica, Universit\`a Cattolica, via Musei 41, 25121
Brescia, Italy}
\affiliation{ I.N.F.N., Sezione di Pavia, Italy}
\author{G.~L.~Celardo}
\affiliation{Dipartimento di
Matematica e Fisica, Universit\`a Cattolica, via Musei 41, 25121
Brescia, Italy}
\author{A.~Musesti}
\affiliation{Dipartimento di
Matematica e Fisica, Universit\`a Cattolica, via Musei 41, 25121
Brescia, Italy}
\author{R.~Trasarti-Battistoni}
\affiliation{Dipartimento di
Matematica e Fisica, Universit\`a Cattolica, via Musei 41, 25121
Brescia, Italy}
\author{P.~Vachal}
\affiliation{Faculty of Nucl. Sci. and Phys. Eng., Czech Technical University,
Prague, Czech Republic}

\begin{abstract}
We demonstrate 
the existence of a topological disconnection threshold, recently 
found in Ref. \cite{JSP}, for generic  $1-d$
anisotropic Heisenberg models
interacting with an inter--particle potential $R^{-\alpha}$ when $0<\alpha < 1$
(here $R$ is the distance among spins).
We also show that if $\alpha $ is greater than the embedding dimension $d$ 
then the ratio between the disconnected  energy region
and the total energy region goes to zero when the number of spins becomes 
very large.
On the other hand, 
numerical simulations in $d=2,3$  for the long-range case $\alpha < d$ support 
the conclusion that such a ratio remains finite for large $N$ values.
The disconnection threshold can thus be thought as a distinctive property of
anisotropic long-range interacting systems.

\end{abstract}
\date{\today}
\maketitle

\section{Introduction}

Despite the wide use in statistical physics, long-range interacting
systems, that is those systems characterized by a
pairwise interaction decaying as a power law of the mutual distance
with an exponent $\alpha$ less than the embedding dimension,
 do not have a well defined thermodynamic limit\cite{rue}.
Also is it not at all clear whether their equilibrium properties can be
described by the ordinary tools of statistical mechanics.
For instance, the nonequivalence between the microcanonical and
the canonical approach has been recently found
in a long-range rotators model in the thermodynamic limit\cite{ruffo}. 

Besides these relevant implications in the foundation of 
statistical mechanics and in theoretical physics as well\cite{gold},
the non-extensive behavior of long range  systems
has nowadays become important for applications too, ranging from 
neural systems\cite{amit} to  spin glasses\cite{ford}. 

Within the class of long-range interacting systems, classical spin
models, widely investigated during the last years\cite{gian},
are the most easy-to-handle both from the analytical and the 
numerical point of view.
Within such class of systems, (to be more precise, a class 
of anisotropic Heisenberg models)
the existence of a threshold of disconnection 
in the energy surface has been demonstrated\cite{JSP}
for an interparticle interaction with infinite range.
It has been called non-ergodicity threshold  
for historical reasons\cite{Palmer}, 
even if the term can  generate some confusion. 
Indeed non-ergodicity is only an obvious
consequence:  it simply means that the energy surface is
topologically disconnected in two regions characterized by
positive and negative magnetization. In other words 
it cannot exist a  dynamical path  connecting them and
all trajectories starting from one region of the phase space
stay there forever.
For this reason we prefer here to call it
Topological Non-connectivity Threshold (TNT). 

The presence of  the TNT  cannot be considered
an exotic  mathematical peculiarity
of some toy model. Its dynamical relevance has been
studied in \cite{firenze}, where an 
explicit expression for the reversal times of
the magnetization (the time necessary to jump from
one branch to the other) has been
given in the neighbors of the critical energy point.
Reversal times diverge at the TNT as a power law with
an exponent dependent on the number of the particles
(and, probably, on the embedding dimension) as in  
ordinary phase transitions. 
Strictly speaking, even if in different context and for different models,
the relationship between energy thresholds and topology transitions
in the configuration space of classical spin models
has been recently investigated\cite{pett}.

Also, while the threshold was explicitly found 
within  a class of anisotropic classical Heisenberg models with an
easy axis of magnetization and all-to-all constant interaction,
at the same time systems with nearest neighbor interaction 
were found to have a different behavior.
For instance,  the 
portion of disconnected energy region grows
with the number of particles $N$, less than 
the energy itself, thus resulting in a zero ratio 
in the thermodynamic limit.
Needless to say, such ratio stays finite for anisotropic coupling and
all to all interaction.

While this feature is surely due to the anisotropy of the coupling 
(such finite ratio disappears for isotropic coupling
even in the case of infinite range interaction),
the question arises whether the presence of the TNT can be considered
a pathological effect of the unphysical infinite interaction
range or it is just somehow related with the long-range effects.
This does not represent an academic question.
Indeed,  despite the possible applications of such model even in the 
case of  all-to-all interaction\cite{cornell},
physical models require taking into account
more realistic  interactions, usually anisotropic\cite{anis} and depending
generically from  the inter--spin distance, as for the dipole-dipole coupling
or when the spin is coupled with the electron spin of the conduction band of
a metal, e.g. the RKKY model \cite{qtm}.
This  leads
quite naturally to  Hamiltonians with an interparticle potential 
decaying as a generic power law with 
an exponent $\alpha$ of the relative distance $R$.
The results found in Ref.\cite{JSP} 
can thus 
be recovered by letting  respectively
$\alpha \to  0$ (all to all coupling)
or  $\alpha \to \infty$ (nearest neighbor coupling).

Here,  we extend the previous  results to the whole class of 
models with an inverse power distance
potential and show that
$\alpha = 1 $ is a critical exponent for non-connectivity
in $d=1$ chains.

In general, the extension to higher dimensions is far from trivial, both
numerically and analytically.  However,  we  prove  that 
the TNT, if any,  can not
``survive'' (and we will specify the precise meaning below)
when  $N\to\infty$, and $\alpha > d$.
Numerical simulations in $2$--d and $3$--d also suggest that, for $\alpha<d$,
the  ratio between the disconnected energy region 
and the total energy range is finite in the thermodynamic limit.
We thus conjecture that the TNT is a generic
property of anisotropic long-range systems in any dimension.

\section{The model}

The  Hamiltonian is a simple generalization
of that considered in Ref.~\cite{JSP}, and it is given by: 
\begin{equation}
\label{ham}
H=-\frac{1}{2}  \sum_{j\ne i}^N  c_{|i-j|} ( S_i^y S_j^y
- \eta S_i^x S_j^x ),
\end{equation}
where $\vec S_i=(S_i^x,S_i^y,S_i^z)$ is the spin vector with continuous
components and modulus $1$, $N$ is the number  of  spins, 
 $\eta$ ($0 \le \eta<1$) is an anisotropic coefficient, 
and
$c_{|i-j|} =  |i-j|^{-\alpha}$, with $\alpha >0$.
For definiteness we consider here only the case of an even number $N$ of 
classical spins.

Such kind of models are 
characterized by a minimal and maximal energy $E_{min},E_{max}$,  
and by a finite 
energy range $E_{max}-E_{min}$ that we call energy spectrum (ES).
In order to define properly  the disconnection threshold, 
let us introduce the set ${\cal A}$ of all spin configurations
with a zero projection of the total
magnetization along the $y$-axis:
\begin{equation}
\label{cala}
{\cal A} = \{ {\cal C} (\vec{S}_1,\ldots\vec{S}_N) | 
\ m_y = \sum_{i=1}^N S_i^y = 0 \}.
\end{equation}
The TNT is thus defined as: 
\begin{equation}
E_{dis} = Min_{{\cal C} \in {\cal A}}   [  H  ],
\label{eds}
\end{equation}
and the spin configurations corresponding to $E_{dis}$ will
be indicated as  ${\cal C}_{dis}$.
Here we are mainly interested in all those cases where
the TNT, if any, occupies a significant portion
of the ES in the thermodynamic limit.
For this reason let us define the disconnection ratio:
\begin{equation}
r=\frac{E_{dis}-E_{min}}{|E_{min}|} >0
\label{erre}
\end{equation}
A system will be considered {\it disconnected} only if
$r\to$ {\it const.} $> 0$, when $N\to\infty$.
Note that the definition of $r$ given in Eq.~(\ref{erre}) has
a meaning only for systems with a bounded energy range.

A dynamical  consequence of the TNT   is that below it, a sample
with a given initial magnetization $m_y$, cannot change the
sign of $m_y$  for any time,  since  the constant energy surface is
disconnected in a positive and a negative magnetization regions,
thus no continuous dynamics can bring an isolated  system 
from one region to the other.

Our proof will follow two steps:  in the first part we  find
the  minima of the $x$ and $y$ parts separately.
Then  we will show that the disconnected ratio 
goes to zero for long-range interaction, while it goes
to some finite constant in the short range case.

\section{One dimensional case}

\subsection{TNT, if any,  is  in the XY plane}

Roughly speaking, since Hamiltonian (\ref{ham}) is independent 
of the $z$-component of spins, the minimum
will occur when the spins are as large as possible in (\ref{ham}), 
namely  in the $XY$ plane. 

In order to prove that the configuration ${\cal C}_{dis}$ 
effectively lies in the $XY$ plane, let us assume that it 
has some $S_z$ component different from zero.
For definiteness assume $S_1^z > 0$. It is then possible
to define another configuration ${\cal C}^\prime$ simply 
making  a rotation around the $y$-axis clockwise or counterclockwise
which puts the spin $S_1^z$  onto the plane $XY$. 
The energy difference between these two
configurations can be computed at glance:
\begin{equation}
\Delta E= \eta \sum_{i=2}^N  c_{i-1} S_i^x  \left( \pm
\sqrt{1-(S_1^y)^2} -S_1^x \right).
\label{dez}
\end{equation}
Here the different sign $\pm$ indicates the different way
(clockwise or counterclockwise)
of rotation.
Since $S_1^x = \pm \sqrt{1-(S_1^y)^2-(S_1^z)^2}$,
it is then clear that, according to this sign 
it is always possible to rotate in such a way to have $\Delta E \le 0$.

The same procedure can be applied   $n$ times for any other $S_i^z\ne 0$, so that
we will end with a configuration ${\cal C}^{(n)} \in {\cal A}$
(the rotation does not change the constraint) 
with energy $E^{(n)}\leq E_{dis}$ .
We can therefore consider  the configurations in the $XY$ plane.
This choice has the  main advantage that 
it is sufficient to consider as independent variables the angles 
$\theta_i$ of the $i$-th spin w.r.t. the $x$-axis, thus satisfying automatically
the conditions on the unit spin modulus
: $S_i^x = \cos \theta_i$ and $ S_i^y = \sin \theta_i$.
Therefore we have to minimize the following expression:
\begin{equation}
\label{ham1}
H=\frac{1}{2}  \sum_{j\ne i}  c_{|i-j|} (\eta\cos\theta_i \cos\theta_j
-\sin\theta_i \sin\theta_j )\equiv \eta H_x + H_y,
\end{equation}
under the constraint $\sum_{i=1}^N \sin \theta_i = 0 $.
Since
\begin{equation}
E_{dis} = Min(H | m_y=0) \ge  Min( \eta H_x) + Min(H_y| m_y=0),
\label{boh}
\end{equation}
a lower bound of $E_{dis}$ can be provided finding the minima of
the two terms in 
the r.h.s. of Eq.~(\ref{boh}).
Note that the first term on the right of Eq.(\ref{boh}) does not contain constraints,
indeed we will show in the next section that the absolute minimum
of $H_x$ automatically satisfied the constraint $m_y=0$.

\subsection{Minimum of $H_x$}
\label{accax}

The minimum of $H_x$ due to the overall plus sign ($\eta > 0 $), 
can be obtained  as in a standard  anti-ferromagnetic spin system
with neighbors interaction, that is
disposing alternatively the spins
along the $x$-axis as $+1$ and $-1$.

Indeed,  for any $\alpha$,
let us call the $k-th$ spin component $S_k^x= s$ and rewrite
the energy as follows: 
$$
E_x = s\sum_{j\ne k} c_{|j-k|} S_j^x + 
\frac{1}{2}\sum_{k\ne i\ne j} c_{|k-i|} S_k^x S_i^x
\equiv a s + b,
$$
where $a,b$ are constants independent  of $s$. 
There are two possibilities, $a=0$ or $a\ne 0$.
In the first case the energy $E_x$ turns out to be independent 
of the $k-th$ spin, while
for $a\ne 0$, 
the minimum is attained when $s$ has its maximal value ($+1$ if
$a<0$, $-1$ if $a>0$). So, in any case we can say that the minimum occurs
when $|s|= 1$. Since the procedure can be iterated for
all spins components, the
minimum  occurs when $S_k^x = \pm 1$ for any $k$, namely 
in the class  of Ising models  $(\sigma_1,...,\sigma_N)$ with $\sigma_i=\pm 1$
and long range interaction.

We still have to prove that the minimal energy is obtained 
when the spins have alternating signs.
To this end, let us consider
the interaction between two neighbor
spin pairs, the $j$-th, namely $\sigma_{2j-1}\sigma_{2j}$ and the
$(j+k+1)$-th, namely $\sigma_{2j+2k+1}\sigma_{2j+2k+2}$ (this can be done 
since there is an even number of spins). There are  
16 possibilities but only six of them have different energy
due to the symmetry on the whole change of sign. They are:
\begin{eqnarray}
\nonumber  &++ \quad ++ \ &E_1=2+c_{2k+1}+2c_{2k+2}+c_{2k+3}\\
\nonumber  &++ \quad +- \ &E_2=c_{2k+1}-c_{2k+3}\\
\nonumber  &++ \quad -+ \ &E_3=-E_2\\
\nonumber  &++ \quad -- \ &E_4=2-c_{2k+1}-2c_{2k+2}-c_{2k+3}\\
\nonumber  &+- \quad +- \ &E_5=-2-c_{2k+1}+2c_{2k+2}-c_{2k+3}\\
\nonumber  &+- \quad -+ \ &E_6=-2+c_{2k+1}-2c_{2k+2}+c_{2k+3}.
\end{eqnarray}

From the monotonicity of the function
$c_x$ one gets,
$$E_5 < E_1,E_2,E_3,E_4.$$
Imposing $E_5 < E_6$ we obtain
$$2c_{2k+2}< c_{2k+1}+c_{2k+3},$$
that holds from the convexity property of $c_x$.

Since $E_5$ is minimal for each pair interaction, 
the absolute minimum will be obtained
using, for each pair, the configuration 
\begin{equation}
\label{ccix}
{\cal C}_x = \{ S^x_i = (-1)^i \}_{i=1}^N. 
\end{equation}
The energy $E_x$ for  this configuration can be computed immediately:
\begin{equation}
E_x = \sum_{k=1}^{N-1} (-1)^k \frac{(N-k)}{k^\alpha}.
\label{edix}
\end{equation}

Let us notice  that such a result is far from being obvious.
Indeed a decreasing non-convex
function $c_x$ could give rise to  a different  minimal configuration. 

\subsection{Minimum of $H_y$} 
\label{accay}

\begin{figure}
\includegraphics[scale=0.34]{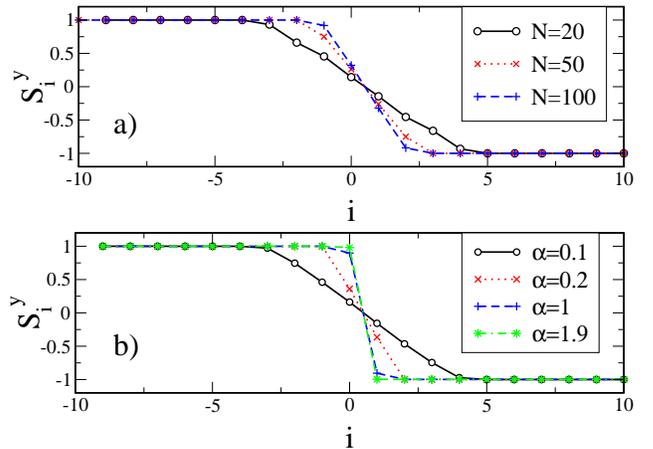}
\caption{(Color online)
Spin values  along the chain {\it vs} the spin index (only the central
part of the chain has been shown)
for the Hamiltonian $H_y$.
a) for fixed $\alpha=0.1$ and different $N$ values as
indicated in the legend;
b) for fixed $N=20$ and different $\alpha $ values (see  the legend).
}
\label{af}
\end{figure}
Let us now switch to the more difficult task (due to the constraint) of
computing $E_y = Min(H_y | m_y=0)$. 
Physically, due to the overall minus sign in front of $H_y$,
one can expect that clusters
of aligned spin with unit modulus 
(ferromagnetic order)
will decrease the energy  with respect to 
other configurations. 
This is surely true for nearest neighbor interaction ($\alpha=\infty$)
 but it can not be true for all $\alpha$ values.
For instance, when $\alpha=0$,  the energy corresponding to the configuration
with half spins equal to $1$ and half equal to $-1$ (the order is irrelevant)
is $E_0=N/2>0$, while the true minimum $E=0$ is attained when all spins are $0$.

Then, the question arises of what can be the minimum in presence of
a generic $\alpha$.

Applying the standard  Lagrange multipliers formalism,
one has to minimize the function:
\begin{equation}
\label{ham44}
H=-\frac{1}{2}  \sum_{j\ne i}  c_{|i-j|} \sin\theta_i \sin\theta_j
-\lambda \sum_i \sin\theta_i.
\end{equation}
where $\lambda$ is the Lagrange multiplier associated to $m_y=0$.

Taking the derivatives, we get, for each spin,  two possible solutions:
\begin{eqnarray}
\label{lagI} & \cos \theta_i = 0, \\ 
\label{lagII} & \sum_{j\ne i}  c_{|i-j|}  \sin \theta_j -\lambda =0.
\end{eqnarray}
However,  solving, even numerically, 
the 
system (\ref{lagI},\ref{lagII})
is more difficult than finding  directly the minimum. 

We have therefore calculated the minimal configuration under 
constraint, using an iterative optimization approach based on 
the FFSQP solver \cite{ffsqp}
and also developing the following approach outlined here below.

\begin{itemize}
\item Start with a random configuration with $m_y=0$.
\item Chose for  the $k-th$ spin a new value between  $-1 $ to $1$ and compute
the energy. This generally produces a change in magnetization 
$\Delta m_y \ne 0$.
\item Distribute equally $\Delta m_y \ne 0$ among the other spins 
taking into account the constraint about their modulus. 
Specifically subtract/add  to every spin the minimum of
its distance from the values $\pm 1$ and the 
mean  of $\Delta m_y$.
\item Iterate over all  spins up  to an energy variation  less than some fixed value 
(from $10^{-3}$ to $10^{-8}$ in our simulations).
\end{itemize}

The two approaches give the same result:
for any initial random configuration the algorithm described above 
converges to some smooth configuration, for any finite $N$ and $\alpha > 0 $,
as indicated in Fig.~\ref{af}.
There,  we considered respectively the case of $\alpha$
fixed varying $N$ (Fig.~\ref{af}a), and   
$N$ fixed varying $\alpha$ (Fig.~\ref{af}b).
Within the numerical errors the spins in the minimal energy
configuration are distributed monotonically
and anti-symmetrically w.r.t. the center of the chain.
Then we  assume  
that $E_y$ is given by an anti-symmetric
distribution of the spin with a non-decreasing (or non increasing)
monotonic dependence 
of the $y-$ spin component along the chain.

An interesting  feature is the presence 
of a finite domain wall (defined by those
spins having length less than $1$) between two clusters  
with $S_i^y=+1$ ($\uparrow$) and $S_i^y=-1$ ($\downarrow$) respectively.
With decreasing range
of interaction (increasing $\alpha$: Fig.~\ref{af}b) or
increasing number of spins (Fig.~\ref{af}a),
the interface region between the clusters
($\uparrow$) and ($\downarrow$) decreases.
This agrees, at least qualitatively,  with the results obtained for 
the nearest neighbor model ($\alpha=\infty$) where the minimal
configuration is given by
$${\cal C}_{\uparrow\downarrow}=
(\uparrow...\uparrow ;\downarrow...\downarrow).$$

Thus, due to long range interaction,  an interface region
between the two clusters with opposite magnetization is produced.
It is, of course,  physically relevant to understand if the size of the interface region
goes to zero in the $N \to \infty$ limit.

Strictly speaking the configuration $E_{\uparrow\downarrow}$ is not
an absolute minimum for any  $\alpha > 0 $ and finite $N$. 
In order to prove that,
consider the configuration
$${\cal C}_s = (\uparrow...\uparrow ;s ; -s ;\downarrow...\downarrow),$$
with
$N-2 $ spins satisfying condition (\ref{lagI}) and the two central ones
satisfying the condition (\ref{lagII}). The energy $E_s$ 
corresponding to ${\cal C}_s$  can be written as
\begin{equation}
E_s = \bar{E}  +c_1 s^2 -2s [ c_1 - c_{N/2}],
\label{edis}
\end{equation}
where $\bar{E}$ is independent of $s$. 
The minimum is thus obtained when  $s= 1-c_{N/2}/c_1 \ne 1$.
The energy difference to $E_{\uparrow \downarrow}$ is
\begin{equation}
\Delta E = E_s - E_{\uparrow\downarrow} = -\frac{c_{N/2}^2}{c_1} = -
\left(\frac{2}{N} \right)^{2\alpha}
<0.
\label{difdie}
\end{equation}
Therefore, for any finite chain and finite $\alpha$, 
$E_s < E_{\uparrow\downarrow}$.

Physically, ${\cal C}_s$ 
has an energy less than ${\cal C}_{\uparrow\downarrow}$
due to border effects.
Indeed  the energy of  two opposite spins of length $1$  is  $E = c_1$, while it is only
$c_1s^2$ for two  shorter spins $|s| <1$. 
On the other hand the interaction between $s$ ($-s$) and the other
spins of length $1$ ($-1$) is canceled  one to one  but 
the interaction 
with the closest spin ($-s \ c_1$) and  with the last
opposite one ($s \ c_{N/2}$).
This gives Eq.(\ref{edis}).

\begin{figure}
\includegraphics[scale=0.31]{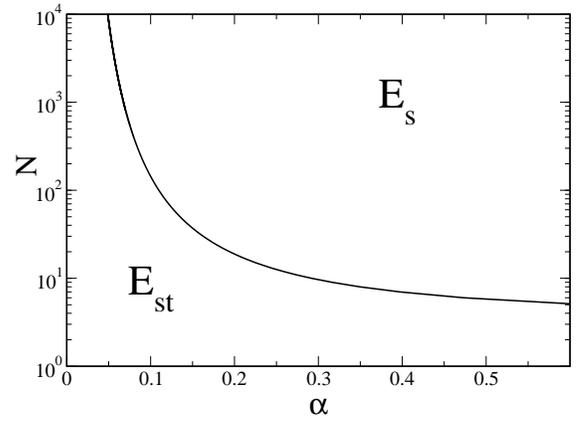}
\caption{Critical $N_{cr}$ as a function of $\alpha$.
The region above the line  is where the TNT is given by $E_s$ (one spin pair
decreased) and the region below it is  where the TNT has more than one
spin pair decreased ($E_{st}$).
}
\label{ncr}
\end{figure}

The same procedure can be applied taking a trial configuration
with energy $E_{st}$:
$$
{\cal C}_{st} = (\uparrow ... \uparrow ; t;  s;  -s ; -t ; 
\downarrow ...\downarrow).
$$

In this case a minimum  with $s<t<1$ can be found only for 
$N<N_{cr}(\alpha) = (2^{\alpha+1} -1/2)^{1/\alpha}$.
Asymptotically, for large $N$, this implies that for 
$N> 2 C^{1/\alpha}$, where $C=2 e^{-1/4}>1$, the minimal solution 
has energy  $E_s$. 
In Fig.~\ref{ncr} we show the graph of $N_{cr}(\alpha)$, and the two regions in
 the plane $(N,\alpha)$,  where $E_s$ is the minimal solution, and where
another minimal solution with four  (or more) spins with length less than 1 is
possible ($E_{st}$). Since $N_{cr} (\alpha) \to \infty$ 
for  $\alpha \to 0$, for any $\alpha\ne 0 $  a sufficiently
large $N> N_{cr}(\alpha)$ value exists  (thus in the thermodynamic limit)
such that $E_s$ is the minimal solution. Then, for $N> N_{cr}(\alpha)$: 
\begin{equation}
E_y = E_{\uparrow\downarrow} - \left( \frac{2}{N}\right)^{2\alpha},
\label{ediyy}
\end{equation}
where $E_{\uparrow\downarrow}$ can be written in closed form as:
\begin{equation}
E_{\uparrow\downarrow}=
\left( \frac{2}{N} \right)^{\alpha-1} + \sum_{k=1}^{N/2-1} \frac{3k-N}{k^\alpha} +
\frac{N/2-k}{(N/2+k)^\alpha}.
\label{ediy1}
\end{equation}

We thus proved that for $\alpha >0$ and $N> N_{cr}(\alpha)$
\begin{equation}
E_{\uparrow\downarrow} > E_{dis} > \eta E_x + E_y,
\label{disse}
\end{equation}

where the expressions for $E_{\uparrow\downarrow}$, $E_x$, and  $E_y$
are given respectively by Eqs.~(\ref{ediy1}, \ref{edix}, \ref{ediyy}).

\subsection{Thermodynamic limit }

\begin{figure}[t]
\includegraphics[scale=0.32]{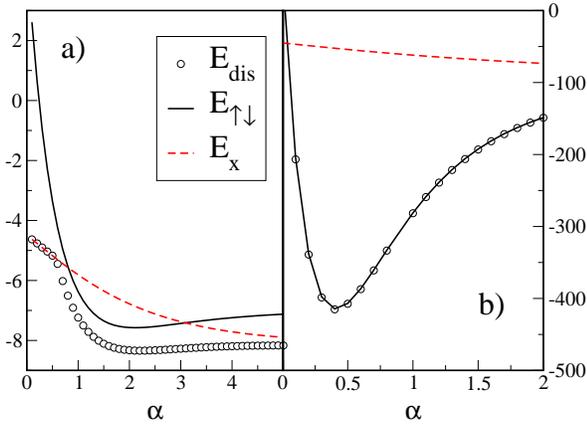}
\caption{(Color online) $E_x$ (full black line),
$E_{\uparrow\downarrow}$ (dashed red line) and $E_{dis}$ (open circles)
versus $\alpha$, for $\eta=0.9$ and different $N$ values
: (a) $N=10$, (b) $N=100$.
}
\label{enexy}
\end{figure}

Let us now show that,
in the long range case  $0<\alpha<1$, 
the ratio $r$ between the disconnected ratio, 
defined by Eq.~(\ref{erre}),  goes to a non zero 
constant when the number of spins
goes to infinity, while, for short range interaction
$\alpha>1$, it goes to zero,
thus revealing the intrinsic long range nature of the TNT.
 
The minimum energy, having as a configuration 
$\ {\cal C}_{min} = \{ S_y^i = 1 \}_{i=1}^N $
(all spins aligned along the y-direction) can 
be easily found:
\begin{equation}
E_{min} = \sum_{k=1}^{N-1} \frac{k-N}{k^\alpha}.
\label{emm}
\end{equation}
Let us also define the quantities:
\begin{eqnarray}
\label{erre1} 
\nonumber r_1 &&= \frac{E_{y}+\eta E_x -E_{min}}{|E_{min}|},\\
r_2 &&= \frac{E_{\uparrow\downarrow}-E_{min}}{|E_{min}|}.
\end{eqnarray}
Due to (\ref{disse}),   $0<r_1<r<r_2$.

\subsubsection{Long Range}
Consider first the long-range case $0<\alpha < 1$. The following asymptotic 
expression, for $N\to\infty$, can be found by substituting sum with integrals:
\begin{eqnarray}
\label{cazemin}
&E_{min} &\simeq -\frac{N^{2-\alpha}}{(2-\alpha)(1-\alpha)}  +O(N),\\
\label{cazsugiu}
&E_{\uparrow\downarrow} &\simeq N^{2-\alpha}\frac{1-2^\alpha}{(1-\alpha)(2-\alpha)}  +O(N),\\
\label{cazex}
& E_x  &\simeq -b_\alpha N +O(N^{1-\alpha}),
\end{eqnarray}
where $b_\alpha > 0 $ is a constant independent of $N$.

Since both $r_1 \to |2-2^{\alpha}|$ and
$r_2 \to |2-2^{\alpha}|$ for $N\to \infty$, it follows
$r\to |2-2^{\alpha}|$ too,
so that the  disconnected energy region remains finite w.r.t. the ES
in the thermodynamic limit. This  prove the disconnection of the
system below the TNT.
It is interesting to note that, as $\alpha \to 1$, $r \to 0$.

\subsubsection{Short Range}
In the short-range case, $\alpha >1$, one can write the following
asymptotic expression (by substituting sums with integrals) :
\begin{equation}
\label{cazemin1}
E_{min} \simeq c_\alpha N + O(N^{2-\alpha}),
\end{equation}
where  $ -1 +1/(1-\alpha) < c_\alpha < -1 +2^{1-\alpha}/(1-\alpha)$.

Let us first show that, as  $N\to \infty$:
\begin{equation}
\label{limi}
\lim_{N\to\infty} \frac {E_{\uparrow\downarrow} -E_{min}}{N} = 0.
\end{equation}
Computing explicitly the l.h.s. of (\ref{limi}) one gets :
\begin{eqnarray}
\label{limi1}
\nonumber&&0 \le \lim_{N\to\infty} \frac{2}{N} \left( \sum_{k=1}^{N/2-1} \frac{1}{k^{\alpha-1}} 
+\sum_{k=N/2+1}^{N-1} \frac{N-k}{k^{\alpha}} \right) \\
\nonumber &&\le \lim_{N\to\infty} \frac{2}{N}\left( \int_1^{N/2} \ dx \ x^{1-\alpha} +\right.\\
&&\left. \int_{N/2+1}^N \ dx \ \frac{N-x}{x^\alpha}\right) = 0.
\end{eqnarray}

Then, $r_2 \to 0 $ and, since $r_2 > r > 0$ it follows $r\to 0$
and the system is not disconnected.

This concludes our proof, which is valid for anti-symmetric coupling
(positive $\eta$).

\subsection{1d, numerical solution for the full model}

Despite the proof of the existence of the TNT did not require 
the explicit knowledge of the spin configuration, it may be of some interest 
to find it.

\begin{figure}[t]
\includegraphics[scale=0.34]{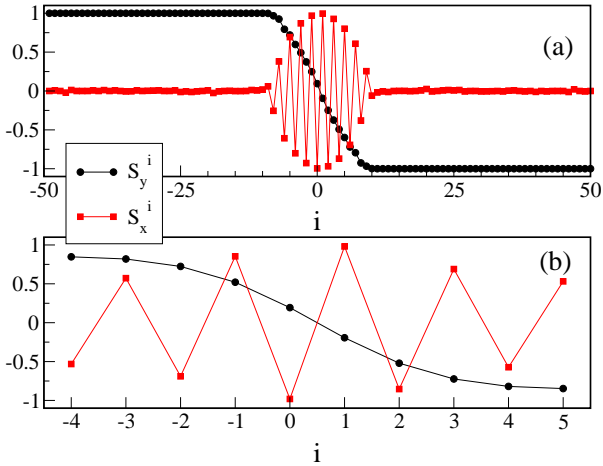}
\caption{(Color online)
(a) $x$ and $y$ spin components for the numerical TNT. Here is $N=100$,
$\eta=0.9$ and $\alpha=0.05$.
(b) $x$ and $y$ spin components for the numerical TNT. Here is $N=10$,
$\eta=0.9$ and $\alpha=\infty$.
}
\label{lastb}
\end{figure}

Finding analytically the spin configuration of the full model 
under the constraint $m_y=0$ for any $\alpha, \eta$ and $N$
is a complicated task.
Indeed, depending on the different values of the parameters, the minimal 
configuration can completely change its shape, 
for instance from all spins along the $x$-axis with alternating signs 
(giving rise to the energy $E_x$)
to  all spins along  the $y$-axis  (first half positive,  second 
half negative) giving rise to $E_{\uparrow\downarrow}$.
For instance, when $\alpha=0$,  
$E_x <0 < E_{\uparrow\downarrow}$, while for $\alpha \to \infty$ 
and $N$ sufficiently large
$E_{\uparrow\downarrow} < E_x < 0 $
(for small $N$  it is also possible to have $E_x<E_{\uparrow\downarrow}<0$).

This is explicitly shown in Fig.~\ref{enexy}, where $E_x$,
$E_{\uparrow\downarrow}$ and $E_{dis}$ (the TNT)  obtained numerically
for two different $N$ values have been plotted as a function of
$\alpha$. 
As one can see, for $\alpha$ less than some value
depending on $\eta$ and $N$, say
$\alpha_0(N,\eta)$,  one has $E_x < E_{\uparrow\downarrow}$, while
for $\alpha > \alpha_0(N,\eta)$, one can have 
different possible situations depending on the $\eta$ and $N$ values. For instance 
for $N=10$ and $\eta=0.9$, $E_x <E_{\uparrow\downarrow}$ (Fig.~\ref{enexy}a) for $\alpha \to \infty$, while 
$E_{\uparrow\downarrow} < E_x$ for $\alpha \to \infty$ and $N=100$ (Fig.~\ref{enexy}b).

It is also instructive to describe the behavior of  $E_{dis}$
as a function of $\alpha$. As one can see (Fig.~\ref{enexy}a), for relatively small
$\alpha$, $E_{dis}$ closely follows $E_x$ while 
for large $\alpha$ values,  $E_{dis}$, even if different from both,  
is closer to $E_{\uparrow\downarrow}$ than to $E_x$.
This is the rule, at least for large $N$ values, and the difference
between the configurations given by $E_{\uparrow\downarrow}$ and $E_{dis}$ is 
only restricted to a small domain wall in the central part of the chain, 
see Fig.~\ref{lastb}a, where the configuration ${\cal C}_{dis}$  in a long range
case has been shown.
Completely different is the situation for small $N$ values, e.g. Fig.~\ref{lastb}b.
Here only 10 particles are considered. In this case, for $\alpha\to\infty$ (see discussion
above) $E_x < E_{\uparrow\downarrow}$ and the configuration  ${\cal C}_{dis}$
is something between ${\cal C}_{x}$ and ${\cal C}_{\uparrow\downarrow}$ 
(see Fig.~\ref{lastb}b.)

\begin{figure}[t]
\includegraphics[scale=0.31]{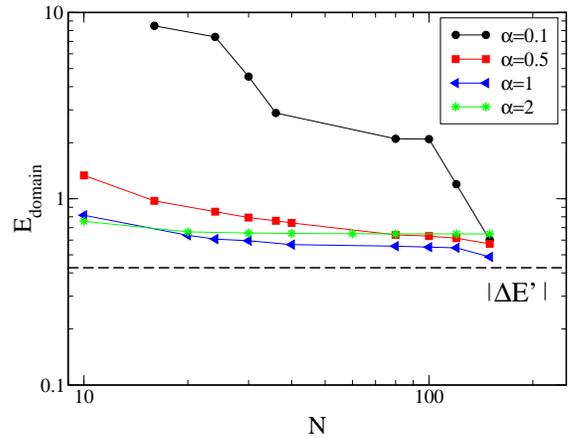}
\caption{(Color online)
Domain wall energy as a function of $N$ for different $\alpha$ values,
as indicated in the legend, and $\eta=0.9$.
Also shown as horizontal dashed line $|\Delta E^\prime|= \eta^2/(1+\eta)$.
}
\label{lasta}
\end{figure}

Let us analyze in detail the behavior of the domain wall under a change
in the parameters of the system.
To this end we define the energy domain as
\begin{equation}
E_{domain} = |E_{dis}-E_{\uparrow\downarrow}|.
\label{domain}
\end{equation}
Its behavior for different $N$ and $\alpha$ values has been shown in
Fig.~\ref{lasta}.
As one can see, in the large $N$ limit, the energy domain approaches,
for any $\alpha$, some finite nonzero value dependent only of $\eta$.
This is remarkably different from the domain wall obtained by minimizing $H_y$
under the constraint $m_y=0$, see Sec.~\ref{accay}, where the formation
of the domain wall was essentially due to border effects and whose energy
disappears for large $N$ values, see Eq.~(\ref{ediyy}).

This asymptotic value can be understood as follows: consider the trial
configuration
\begin{equation}
\label{cxy}
{\cal C}_{xy} =
\left\{ \begin{array}{lll}
S_x^i &=& \{  0,\ldots,0,+\sqrt{1-s^2}, -\sqrt{1-s^2},0,\ldots,0\} \\
S_y^i &=& \{1,\ldots,1,+s, -s,-1,\ldots,-1\}. 
\end{array}\right.
\end{equation}
The energy $E^{xy}(s)$ of this configuration is  given by
\begin{equation}
E^{xy}(s) = \bar{E}  +c_1 s^2 -2s [ c_1 - c_{N/2}]-\eta c_1 (1-s^2),
\label{edis1}
\end{equation}
where $\bar{E}$ is independent of $s$ and $\eta$.

The minimum, as a function of $s$, is
\begin{equation}
s_{min} = \frac{c_1-c_{N/2}}{c_1(1+\eta)}.
\label{mids}
\end{equation}
Then,  for $N\to\infty$, $s_{min}\to 1/(1+\eta) $, which is
independent from $\alpha$. This value can be compared with our numerical results.
The energy difference to $E_{\uparrow\downarrow}$
in the limit $N\to\infty$ is:
\begin{equation}
\Delta E^\prime = E^{xy}(s_{min})- E_{\uparrow\downarrow}  = -\frac{\eta^2}{1+\eta} < 0.
\label{emids}
\end{equation}
Its absolute value has been indicated  as a 
horizontal dashed line in Fig.~\ref{lasta}. As one can see, all curves 
are close to  $|\Delta E^\prime |$ even at $N \sim 100$. 
While we cannot exclude that other configurations, with four or more central spins
$S_i^y<1$, have an energy less than $E^{xy}(s_{min})$, we surely have found a minimal
configuration whose energy is differing from $E_{\uparrow\downarrow}$ 
for a finite quantity in the limit $N\to\infty$.
This in turn implies that the domain does not disappear in the thermodynamic limit.

\begin{figure}[t]
\includegraphics[scale=0.31]{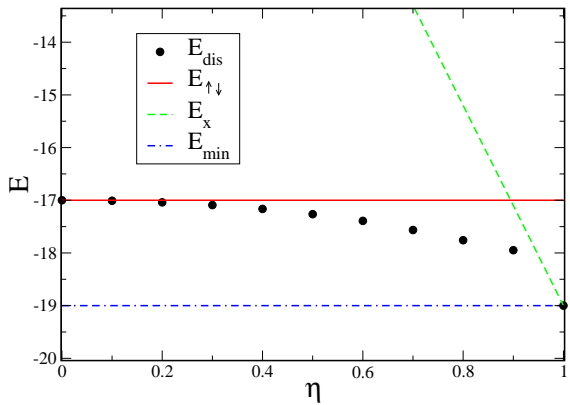}
\caption{(Color online)
TNT for the nearest neighbor interaction ($\alpha=\infty$) {\it vs} the
parameter $\eta$.
Here is $N=20$. Different energies are indicated in the caption inside the figure.
}
\label{n20}
\end{figure}

As a last remark, let us stress that the  no-disconnection
of a system does not necessarily imply the
non existence of a TNT different from the minimal energy at finite $N$,
even for short range interaction. To this end we consider the strongest
short range coupling, namely the nearest neighbor one ($\alpha=\infty$)
and compute numerically
the TNT. Results are presented in Fig.~(\ref{n20}).

As one can see the numerically computed $E_{dis}$ is different from 
$E_{min}$ for  $\eta \ne 1 $ so that a finite range 
of energies $E_{min} < E < E_{dis}$ exists for finite $N$ and nearest neighbor
interaction. Increasing $N$, the size of this energy range remains constant, while
$E_{min} \sim N$. That is why the ratio $r\to 0 $ for large $N$ values.
From the same Fig.~\ref{n20} it is clear that ${\cal C}_{dis}$ goes
continuously from a configuration close to ${\cal C}_{\uparrow\downarrow}$
when $\eta\ll 1$ to one close to ${\cal C}_{x}$  when $\eta \simeq 1$.

\section{Multidimensional case}
\label{sec:multi}

\begin{figure}[t]
\includegraphics[scale=0.34]{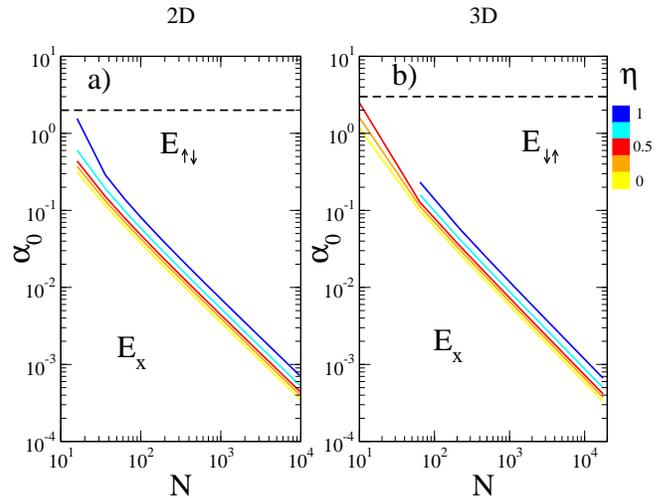}
\caption{(Color online) Critical $\alpha_0$ as a function of
the number of spins
for different $\eta$ indicated by different colors.
(a) $d=2$, (b) $d=3$.
}
\label{cri}
\end{figure}

The results obtained in the previous Sections for $d=1$ can 
be extended in greater dimension $d\ge 2$.

While it can be easily shown that, in the short--range
case $d<\alpha$, the system cannot be disconnected
in the thermodynamic limit,
the proof of the disconnection for the long--range case
is essentially based on the assumption that the minimum
energy with the constraint $m_y=0$ is given by an obvious
extension of what we have found in $d=1$. This is, at the moment,
justified only by our numerical simulations.

Let us then consider a $d$--dimensional  hypercube  
of side $L$, such as $L^d = N$ and
divide it in two equal halves.
Let us then put half of the spins with $y$-component in one region
and the other half in the remaining with opposite $y$-component
and  call $E_{\uparrow\downarrow}$ 
the resulting energy for such configuration. Surely the TNT has
an energy value less or equal to $E_{\uparrow\downarrow}$, that is
$E_{dis} \le E_{\uparrow\downarrow}$.

Let us then write:
\begin{eqnarray}
\nonumber E_{min} &= E_{\uparrow} + E_{\uparrow} + V_{\uparrow\uparrow},\\
E_{\uparrow\downarrow} &= E_{\uparrow}+E_{\downarrow}+
V_{\uparrow\downarrow},
\label{eppp}
\end{eqnarray}

where $E_{\uparrow}, E_{\downarrow}$ are the energies of the
respective halves  
and $V_{\uparrow\downarrow}, V_{\uparrow\uparrow}$ are  the interaction energies between
the two halves with, respectively,  antiparallel  and parallel spins.

Since $E_{\uparrow}=E_{\downarrow}$ and $ -V_{\uparrow\uparrow}=
V_{\uparrow\downarrow}> 0 $, one has:
\begin{equation}
\label{diddd}
0 \le r \le r_{\uparrow\downarrow} =2\frac{V_{\uparrow\downarrow}}{|E_{min}|}
= 2\frac{2E_{\uparrow}- E_{min}}{|E_{min}|}.
\end{equation}

We will make use of the results found in Ref.\cite{thebibble},
that in our variables read as:

\begin{equation}
\label{tama}
\lim_{N\to\infty} \frac{E_{min} (d,\alpha, N)}{N^{2-\alpha/d} - N} = C_d (\alpha),
\end{equation}
for $\alpha \ne d$ and where the constant $C_d (\alpha) >0 $ for $d<\alpha$ 
and $C_d (\alpha) <0 $ for $d>\alpha$
depends only on $d$ and $\alpha$.

\subsubsection{Short-Range}
Let us discuss the short range case $\alpha > d$.
In this case we have,
\begin{equation}
\label{tama1}
\lim_{N\to\infty} \frac{E_{min} (d,\alpha, N)}{N} = C_d (\alpha),
\end{equation}
and, since $E_{\uparrow} = E_{min} (d, \alpha, N/2)$, we can write 

\begin{equation}
\label{didd1}
0 \le r \le r_{\uparrow\downarrow} 
=2 \frac{2E_{\uparrow}/N- E_{min}/N}{|E_{min}/N|} \to 0 ,\ {\rm for} \ N\to\infty.
\end{equation}

This proves that, in the short--range case, $r\to 0 $  for $N\to \infty$.

\subsubsection{Long-Range}

In the long range-case, $\alpha<d$, let us assume 
that, for large $N$ values, $E_{dis} \to E_{\uparrow\downarrow}$.

Estimate (\ref{tama}) becomes in this case:
\begin{equation}
\label{tama2}
\lim_{N\to\infty} \frac{E_{min} (d,\alpha, N)}{N^{2-\alpha/d}} = C_d (\alpha),
\end{equation}

so that, for $N\to\infty$,

\begin{equation}
\label{didd2}
 r \simeq   r_{\uparrow\downarrow} 
= 2 \frac{2E_{\uparrow}/N^{2-\alpha/d}- E_{min}/N^{2-\alpha/d}}{|E_{min}/N^{2-\alpha/d}|} \to 
2-2^{\alpha/d}.
\end{equation}

That way, $r \to const \ne 0 $ for $N\to \infty$ and $\alpha\ne d$, and a finite disconnected
energy range exists in the thermodynamic limit.

It is also interesting to note that, as $\alpha \to d $, $r\to 0$, so that the result 
(\ref{didd1}) is recovered.

The disconnection of the system  in the long range case 
can thus be proved   if we assume  ${\cal C}_{dis} \sim {\cal C}_{\uparrow\downarrow}$.

\begin{figure}[t]
\includegraphics[scale=0.7]{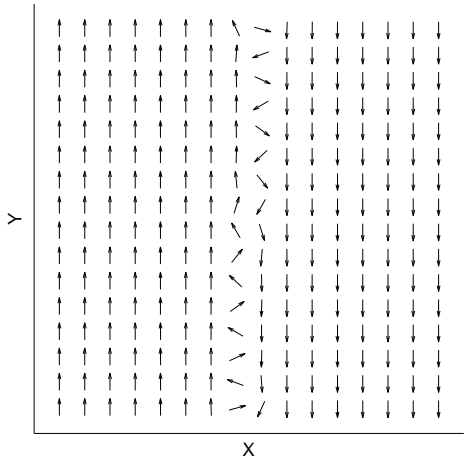}
\caption{(Color online) TNT for the 2-d square lattice, for the
case $\alpha=0.1$,
$\eta = 0.5$, $N=16 \times 16 = 256$ spins.  
Here is $E_{dis}=-655.968997$, while $E_{\uparrow\downarrow}=-654.7308$.
}
\label{sx}
\end{figure}

Numerical simulations confirm this assumption.
Indeed, let us define $\alpha_0(\eta,N)$ as the smallest value such that, 
$$E_x(\alpha_0,\eta,N)  = E_{\uparrow\downarrow}(\alpha_0,\eta,N).$$
Its general dependence on parameters has been presented in Fig.~\ref{cri}
for $d=2,3$.
As one can see, $\alpha_0 \sim 1/N \to 0$ when $N\to\infty$.
In the same picture we indicate  the regions where $E_x$
or $E_{\uparrow\downarrow}$ are respectively the minimal energies
satisfying the constraint $m_y=0$.
Also, $\alpha_0=d$ is plotted as a dashed horizontal line,
showing that the short-range case (above the
line) is characterized by $E_{\uparrow\downarrow}$, while the long-range
case (below the line) can have different behaviors 
($E_x$ or $E_{\uparrow\downarrow}$), even if physically interesting
long-range interactions are generally characterized by $E_{\uparrow\downarrow}$.

As for the spin configuration ${\cal C}_{dis}$,
both for $d=2$ and $d=3$, all physically significant 
cases can be represented by ${\cal C}_{\uparrow\downarrow}$.
Deviations occur for $\alpha << d$, where a domain wall 
appears, see for instance Figs.~\ref{sx}.
As one can see, ${\cal C}_{dis}$
is generically represented by two macroscopic blocks,
with opposite sign of the $y$-magnetization, with  a  domain
wall at their interface.
In the domain wall
the $y$-components increase in absolute value toward the center
and  the $x$-components
are more or less distributed with alternating signs, 
see Figs.~\ref{sx}.

\begin{figure}[t]
\includegraphics[scale=0.31]{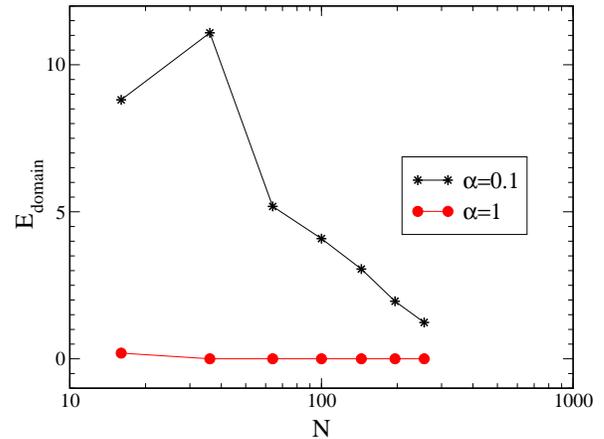}
\caption{(Color online) Domain energy  as a function of the number of lattice spins, for the 2-d square lattice
and $\eta=0.5$, asterisks ($\alpha=0.1$), circles ($\alpha=1$).
For $\alpha >1$, $E_{domain}$ becomes smaller than the computer precision.
}
\label{pav2}
\end{figure}

Defining the domain energy, as the
difference between the numerically found TNT and the
energy $E_{\uparrow\downarrow}$ (\ref{domain}), one has that 
with increasing $N$ it goes to some constant or zero value
(see Fig.~\ref{pav2})  so that,
in the thermodynamic limit 
$E_{dis} \sim E_{\uparrow\downarrow}$, which justify, at least
numerically, our previous assumption.
This concludes our proof of  the disconnection of the system with $\eta \ge 0$
for long--range interaction  in any dimension $d$.

\section{Negative $\eta$}

In this last part we briefly discuss the case $\eta < 0$. 
First of all  $\eta >0$
is not a necessary condition for the existence of a finite
disconnection region. 

Let us first consider the  $1$--d case; In Eq.~(\ref{boh})
$\eta H_x$ becomes ferromagnetic  as $H_y$,  and
the configuration
\begin{equation}
\label{ccxx}
{\cal C}_{x}^\prime = \{ S_x^i= 1\}_{i=1}^N \in {\cal A},
\end{equation}
has an energy $E_x^\prime < E_x$
(which is the energy of the configuration 
${\cal C}_x$, see Eq.~(\ref{ccix})).
Moreover, since the number of parallel spins in 
${\cal C}_{x}^\prime$  
is larger than  in  ${\cal C}_{\uparrow\downarrow}$,
we will expect that the energy $E_x^\prime$,
even if decreased by a factor $\eta$, will become
soon or later less than  $E_{\uparrow\downarrow}$.

While this has no consequences in the
case $\alpha >1$, (we still have $r_2\to 0$ and the 
TNT does not exist), in the
long-range case ($0<\alpha<1$) some interesting features appear.

From Eq.~(\ref{erre1})  one has $r_1 \to |2+\eta-2^\alpha|/2$
for $N\to\infty$ and
a finite disconnection energy region  still exists
for $2^{\alpha}-2<\eta<0$.

\begin{figure}
\includegraphics[scale=0.34]{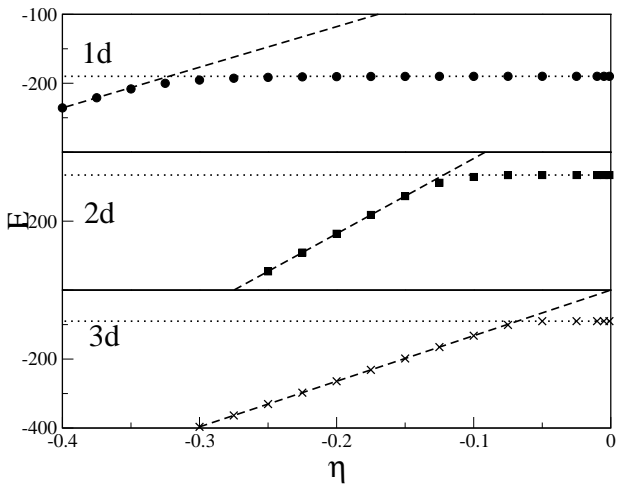}
\caption{
$E_{dis}$ (symbols) as a function of $\eta$ for the long range case $\alpha=0.5$
and $N=64$ ($d=1$), $N=8 \times 8 $ ($d=2$), $N=4 \times 4 \times 4 $ ($d=3$).
Also shown as dotted horizontal lines $E_{\uparrow\downarrow}$,
and, as dashed transverse lines $E_x^\prime$.
}
\label{etaneg}
\end{figure}

In the other case $-1<\eta< 2-2^{\alpha}$ nothing can be
said, even if, according to our numerical simulations
${\cal C}_{dis} \sim {\cal C}_{x}^\prime$.
This effectively happens in  all dimensions $d=1,2,3$, as indicated
in Fig.~\ref{etaneg}, where $E_{dis}$ as a function of $\eta$
as been shown  in a  long range case.
As one can see $E_{dis}$ is close to $E_x^\prime$
(equal, within numerical accuracy, according
to our simulations)  for 
$\eta < \eta_{cr} (\alpha, N)$, while it becomes close to
$E_{\uparrow\downarrow}$, for $\eta > \eta_{cr} (\alpha, N)$.
That holds true  in any dimensions.

Let us note that, as a realistic $\eta_{cr}$, we can assume
the intersection point between $E_x^\prime$
and $E_{\uparrow\downarrow}$, see Fig.~(\ref{etaneg}).
An estimate, that holds only in the thermodynamic limit
can be obtained following the considerations made in Sec.(\ref{sec:multi})
giving $\eta_{cr} \sim  1-2^{\frac{\alpha}{d}}$.

It is also  clear that assuming:
$$E_{dis} = \eta E_x = -\eta E_{min}$$ one has
$r\to 1+\eta$, and the system is disconnected
even for negative $\eta$ in all dimensions.

\section{Conclusions}

Summarizing,  we have studied the occurrence of a topological
non--connectivity threshold (TNT)
in  anisotropic Heisenberg models in $d=1,2,3$ with
an interaction strength depending on a power law
of their relative distance with the exponent $\alpha$.
We have found that the system, in the thermodynamic limit, is disconnected 
only in presence of a long-range interaction $0 \leq \alpha < d$.
On the other side,  in the short-range case, the 
ratio between the disconnected energy region
and the total energy region goes to zero when $N\to\infty$.
The anisotropy represents in this class of systems a necessary condition :
indeed, in the isotropic case, the TNT coincides with the
minimal energy, thus there is no disconnected energy region.

Future investigations concern the experimental evidence of TNT, 
for instance by looking for the divergence of de-magnetization 
times\cite{firenze} 
as a function of energy in small magnetic samples.

Finally, let us point out that from a quantum mechanical point of 
view the classical disconnection does not exclude the flipping
of the magnetization  through Macroscopic Quantum Tunneling
\cite{chud}.
Thus the existence of TNT could give the possibility
to study the emergence of Macroscopic Quantum Phenomena 
in a wide energy range (for macroscopic long range interacting systems),
as has been shown in \cite{quant}, where the quantum signatures
of the TNT in an anisotropic Heisenberg model with all-to-all
interaction have been studied, and the relevance of the TNT
w.r.t. Macroscopic Quantum Phenomena addressed.

\section{Acknowledgment}
We acknowledge useful discussion with J.Barre,
F.M.Izrailev, R.Loubere , D.Mukamel and S.Ruffo.



\begin{thebibliography}{}

\bibitem{JSP} F.~Borgonovi, G.~L.~Celardo, M.~Maianti, E.~Pedersoli,
J. Stat. Phys., {\bf 116},  1435 (2004)


\bibitem{rue} D.~Ruelle, Helvetica Physica Acta, {\bf 36}, 183 (1963).

\bibitem{ruffo} J.~Barr\'e, D.~Mukamel, S.~Ruffo,  Phys. Rev. Lett. {\bf 87}, 3, (2001)

\bibitem{gold} N.~Goldenfeld, {\it Lectures on Phase Transitions
and the Renormalization Group} Addison-Wesley, Reading, MA (1992).


\bibitem{amit} D.J.~Amit {\it Modeling Brain Functions}, Cambridge
University Press, Cambridge, UK (1989).

\bibitem{ford} P.J.~Ford, Contemp. Phys. {\bf 23}, 141 (1982).

\bibitem{gian} A.~Campa, A.~Giansanti and D.~Moroni,
Phys. Rev. E {\bf 62}, 303 (2000); F.~Tamarit and C.~Anteneodo, Phys. Rev. Lett.
{\bf 84}, 208 (2000).

\bibitem{Palmer} R.G.~Palmer, Adv. in Phys., {\bf 31}, 669 (1982).

\bibitem{firenze} G.L.~Celardo, J.~Barr\'e, F.~Borgonovi and S.~Ruffo, cond-mat/0410119.

\bibitem{pett}  L.~Casetti, M.~Pettini, and E.G.D.~Cohen, 
J. Stat. Phys. {\bf 111}, 1091 (2003).

\bibitem{cornell} L.Q.~English et al., Phys. Rev. B ,{\bf 67}, 24403 (2003);
M.~Sato et al., Jour. of Appl. Phys.,  {\bf 91} , 8676 (2002).

\bibitem{anis}   P.~Bruno, Phys. Rev. B {\bf 39},  865 (1989);
    D.~Weller et al,    Phys. Rev. Lett. {\bf 75}  3752 (1995);
    J.~Dorantes-Davila and G.M.~Pastor, Phys. Rev. Lett.
    {\bf 81}, 208 (1998);
    M.~Pratzer et al, Phys. Rev. Lett. {\bf 87}, 127201 (2001).

\bibitem{qtm}  R.M.~White, "Quantum Theory of magnetism", McGRAW-HILL (1970).

\bibitem{ffsqp} J.L.~Zhou, A.L.~Tits and C.T.~Lawrence,
{\it User's Guide for FFSQP Version 3.7},
Systems Research Center TR-92-107r2, University of Maryland,
College Park, MD (1997).

\bibitem{thebibble} S.A.~Cannas and F.A.~Tamarit, 
Phys. Rev. B, {\bf 54}, R12661 (1996); 
T.~Dauxois, S.~Ruffo, E.~Arimondo, M.~Wilkens Eds.,
Lect. Notes in Phys.,  {\bf 602},  Springer (2002).

\bibitem{chud} E.M.~Chudnovsky and J.~Tejada, {\it Macroscopic Quantum
Tunneling of the Magnetic Moment}, Cambridge University Press,
(1998).

\bibitem{quant} F.~Borgonovi, G.L.~Celardo and G.P.~Berman, cond-mat/0506233.

\end{thebibliography}
\end{document}